\documentclass[preprintnumbers,superscriptaddress,nofootinbib,aps,prd,floatfix]{revtex4}

\usepackage{hyperref}
\usepackage{amsmath,slashed}
\usepackage{graphicx}

\newcommand{\be}{\begin{eqnarray*}}
\newcommand{\ee}{\end{eqnarray*}}

\newcommand{\bee}{\begin{eqnarray}}
\newcommand{\eee}{\end{eqnarray}}
\newcommand{\beeq}{\begin{equation}}
\newcommand{\eeeq}{\end{equation}}

\newcommand{\GeV}{~GeV}

\newcommand{\ifb}{{\rm{fb}}^{-1}}

\newcommand{\pt}{p_{\rm{T}}}

\newcommand{\cp}{${\cal{CP}}$}
\newcommand{\cpeven}{${\cal{CP}}$-even }
\newcommand{\cpodd}{${\cal{CP}}$-odd }

\begin{document}

\title{Constraining \cp-violating Higgs Sectors at the LHC using gluon fusion}

\begin{abstract}
We investigate the constraints that the LHC can set on a 126~GeV Higgs boson that is an admixture of \cp~eigenstates. Traditional analyses rely on Higgs couplings to massive vector bosons, which are suppressed for \cpodd couplings, so that these analyses have limited sensitivity. Instead we focus on Higgs production in gluon fusion, which occurs at the same order in $\alpha_S$ for both \cpeven and odd couplings. We study the Higgs plus two jet final state followed by Higgs decay into a pair of tau leptons. We show that using the 8~TeV dataset it is possible to rule out the pure \cpodd hypothesis in this channel alone at nearly 95\%~C.L, assuming that the Higgs is \cp-even. We also provide projected limits for the 14~TeV LHC run.
\end{abstract}

\author{Matthew J. Dolan}
\affiliation{Theory Group, SLAC National Accelerator Laboratory, Menlo Park, CA 94025, USA} 
\author{Philip Harris}
\affiliation{CERN, CH-1211 Geneva 23, Switzerland }
\author{Martin Jankowiak}
\affiliation{Institut f\"ur Theoretische Physik, Universit\"at Heidelberg, Philosophenweg 16, Heidelberg D-69120, Germany}
\author{Michael Spannowsky}
\affiliation{Institute for Particle Physics Phenomenology, Department
  of Physics,\\Durham University, DH1 3LE, United Kingdom}

\preprint{IPPP/14/57}
\preprint{DCPT/14/114}
\preprint{SLAC-PUB-15985}

\maketitle

\section{Introduction}
\label{sec:intro}

The discovery of the Higgs boson at the Large Hadron Collider (LHC)~\cite{:2012gk,:2012gu} marks the beginning of a long and detailed experimental program to measure and constrain the couplings and quantum numbers of the new resonance. In particular, efforts are underway to measure whether the new particle is even or odd under the \cp~transformation, with current results apparently disfavoring the \cpodd hypothesis by nearly $3\sigma$~\cite{Aad:2013wqa,Chatrchyan:2013mxa}.

However, there are numerous examples of extensions of the Standard Model Higgs sector where \cp~is violated and is not a good quantum number of the Higgs-like state (see~\cite{Accomando:2006ga} for a review of a large number of such scenarios). In these models, and indeed in general, one is interested in constraining the properties of the admixture and the extent to which the Higgs is \cpeven or odd, rather than asking whether it is 
100\% one or the other. The discovery that the Higgs has a non-trivial \cp~coupling structure would be direct evidence for Beyond the Standard Model (BSM) physics with many important implications, for instance in baryogenesis~\cite{Bernreuther:2002uj}. While it is known that such studies are difficult, given the continually advancing nature of reconstruction and statistical techniques it is worth investigating the prospects for constraining a mixed-\cp~Higgs at the LHC.

Measuring the \cp~eigenvalue of the Higgs (assuming that \cp~is conserved) is a subject with a long pedigree and extensive literature.
Many of the searches and variables proposed to constrain the \cp~properties of the Higgs rely on its couplings to massive vector bosons. Constraints can be set either by exploiting angular correlations between the leptons from the $ZZ^*\to 4l$ or $2l2j$ decays~\cite{Buszello:2002uu,Choi:2002jk,Godbole:2007cn,Hackstein:2010wk, Englert:2010ud} or through angular correlations in the tagging jets in the weak boson fusion (WBF) production mechanism~\cite{Plehn:2001nj,Odagiri:2002nd,Figy:2004pt,Englert:2012xt,Artoisenet:2013puc,Maltoni:2013sma}. In either case, these methods rely on the existence of unsuppressed (tree-level) couplings between the Higgs and the massive vector bosons. 

While this is the case for the \cpeven component of the Higgs, which couples to the massive vector bosons $V=(W,Z)$ through the $h V^\mu V_\mu$ operator, the \cpodd coupling enters at dimension five through the $h V^{\mu\nu} \widetilde{V}_{\mu\nu}$ operator, where $V^{\mu\nu}$ is the field strength operator for $V^\mu$. Accordingly, \cpodd effects in $h\to ZZ^*$ decays and WBF are suppressed by $\mathcal{O}(\alpha_{EW})$, so that these methods effectively project out the \cpodd part of the Higgs (although see~\cite{Chen:2014gka} for a study which incorporates loop effects and~\cite{Bernreuther:2010uw} for a discussion of $h\to VV$ decays in some specific BSM models).

Such studies often assume that BSM physics enters at a low enough scale such that the dimension five operator contributes at the same order of magnitude as the tree-level \cpeven contribution. However, the existence of light electroweakly interacting states necessary for such a large enhancement of the \cpodd couplings to massive vector bosons is now being directly probed by LHC searches for BSM physics, where no signals inconsistent with the SM have been observed. Furthermore, such states would likely lead to large deviations from SM phenomenology in Higgs boson decays to electroweak gauge bosons, which are also in good agreement with the Standard Model.

Instead it is more promising to study the possible \cpodd admixture of the $126$ GeV resonance via interactions  where the \cpeven and \cpodd couplings are induced at the same order. At tree-level this includes the couplings to quarks and leptons and at loop-level the couplings to gluons and photons. One gluon-induced production process where it is known that sensitivity to Higgs \cp~properties is preserved is $pp \to h+2j$~\cite{DelDuca:2001fn}. As in the WBF channel, the main sensitivity is expected to come from angular correlations between the two tagging jets~\cite{Hankele:2006ja,DelDuca:2006hk,Klamke:2007cu,Campanario:2010mi, Campanario:2014oua}, correlations which can also be exploited in diffractive Higgs production at the LHC~\cite{Khoze:2004rc}. Unlike WBF, in this case the \cpeven and \cpodd contributions are of the same order with the relevant operators being $hG^{\mu\nu}G_{\mu\nu}$ and $h G^{\mu\nu}\widetilde{G}_{\mu\nu}$, respectively.

While the Higgs decay mode $h\to \gamma\gamma$ followed by conversion of both photons to $e^+ e^-$ pairs has recently been suggested as a possible final state for probing Higgs \cp~properties~\cite{Voloshin:2012tv,Bishara:2013vya}, we instead elect to utilize the $h \to \tau\tau$ decay mode.  The majority of previous studies on \cp~in $h\to \tau\tau$ focus on methods for measuring the polarization properties of the Higgs decay products~\cite{Bower:2002zx,Berge:2008wi,Berge:2011ij,Berge:2012wm,Berge:2013jra}. This requires knowledge of the impact parameter or rest frame of the $\tau$s, both of which are difficult quantities to reconstruct in a hadron collider environment (although see~\cite{Englert:2012ct,Harnik:2013aja}). 

Any collider study of Higgs \cp~properties must be compared with measurements from other sources. Particularly relevant are measurements of electric dipole moments (EDMs)~\cite{McKeen:2012av,Brod:2013cka}, which lead to very strong constraints on mixing between \cpeven and \cpodd Higgs components. These constraints, however, rely on the existence of SM-strength interactions of the Higgs to electrons, an assumption that cannot be put to the test at the LHC. Constraints from EDM experiments are therefore complementary to the analysis strategy followed here.
Conceivably, we might discover evidence for \cp~violation in gluon fusion, which, together with a null signal from EDM experiments, would reveal invaluable information about Higgs couplings to the first generation. 

We find that using a set of cuts modeled on the current CMS $h\to\tau\tau$ analysis~\cite{Chatrchyan:2014nva} that data from the 8~TeV run of the LHC is already sufficient to exclude a \cpodd Higgs boson at nearly 95\% C.L..  This can be compared with current bounds presented in ref.~\cite{Freitas:2012kw,Djouadi:2013qya}, which reinterpret current data to set limits on Higgs \cp~properties using measured rates for Higgs production and find constraints at a similar level. Note however that arguments based upon rates alone will always have a flat direction due to possible rescalings of the couplings and Higgs width, and so a differential analysis strategy such as ours should be more robust. 

The rest of this paper is organized as follows. In Section~\ref{sec:model} we introduce the parameterization of \cp~violating effects which we will study: the Standard Model Lagrangian augmented with \cp-violating terms and higher dimensional operators encoding the effects of particles running in loops. In Section~\ref{sec:method} we discuss our methodology and simulations. In Section~\ref{sec:results} we present our results for the expected limits from current LHC data and projections for the limits that can be set with the 14~TeV dataset, before presenting our conclusions and possible directions for future research in Section~\ref{sec:conclusion}.

\section{The Model}
\label{sec:model}

There is a wide variety of models in the literature that lead to \cp~violation in the Higgs sector, such as generalized Two-Higgs Doublet Models, the \cp~violating Minimal Supersymmetric Standard Model (often studied in the CPX~\cite{Carena:2000ks} scenario), and other supersymmetric models that involve R-parity violation~\cite{Accomando:2006ga}. Such scenarios involve a rich UV spectrum of states that is the subject of various LHC searches. In this article we wish to be as model independent as possible and so keep only the 126~GeV Higgs as part of the spectrum, assuming that other BSM states are either out of direct reach of the LHC or that their effects are subdominant for this analysis.

Our model consists of the Standard Model but with the Lagrangian augmented in the following way to include \cp-violating couplings. Following~\cite{Klamke:2007cu} we include couplings between Standard Model fermions and the resonance $h$ which we associate with the Higgs boson:
\begin{equation}
\mathcal{L}_{h\bar{f}f} = \cos\alpha \,  y_{f} \bar{\psi}_f\psi_f h + \sin\alpha \, \widetilde{y}_{f} \bar{\psi}_f i \gamma_5 \psi_f h  \, .
\end{equation}
We have introduced a mixing angle $\alpha$ such that $\cos\alpha=1$ (equivalently $\alpha=0$) corresponds to a Standard Model-like \cpeven Higgs, while $\sin\alpha=1$ (equivalently $\alpha=\pi/2$) corresponds to a \cpodd pseudoscalar. This allows us to study the \cp~properties of the resonance $h$ as a continuous function of the mixing angle $\alpha$.
We will also assume that $y_{f}=\widetilde{y}_{f}=m_f /v$. Having fixed the interactions with fermions allows us to derive the dimension five operators that govern the interaction of $h$ with massless vector bosons, obtaining~\cite{Kauffman:1993nv,Spira:1995rr}
\begin{equation}
\label{eqn:lhgg}
\mathcal{L}_{hgg} = \cos\alpha \, \frac{\alpha_S}{12\pi v}  h G_{\mu\nu}^{a} G^{a,\mu\nu} + \sin\alpha \, \frac{\alpha_S}{4\pi v} h G_{\mu\nu}^{a} \widetilde{G}^{a,\mu\nu}
\end{equation}
for the gluonic interactions, where $v$ is the vev of the SM Higgs, and $\widetilde{G}_{\mu\nu}= \frac{1}{2}\epsilon_{\mu\nu\rho\sigma}G^{\rho\sigma}$ is the dual field-strength tensor. Note that when generating events for our analysis we do not integrate out the top quark, keeping its full mass dependence throughout, so that the effective operators in Eqn.~\ref{eqn:lhgg} should be understood as convenient shorthand. 

The leading order contribution to the interactions of the Higgs with the massive vector bosons is given by:
\begin{equation}
\mathcal{L}_{hVV} \supset \cos\alpha \, \frac{2m_{W}^2}{v} h W_\mu W^{\mu}+ \cos\alpha \, \frac{2m_{Z}^2}{v} h Z_\mu Z^{\mu}
\end{equation}
We neglect higher-order terms, which are loop suppressed by $\mathcal{O}\left( \alpha_{EW} \right)$ relative to this, although see~\cite{Bernreuther:2010uw} for a discussion of how large these terms can become in some BSM models. Note that while the SM matter fields also induce dimension five operators that lead to the decay $h\to \gamma\gamma$, they do not play a role in this article.

\section{Method}
\label{sec:method}

 \subsection{Event generation}

We generate signal events at leading order using \texttt{VBFNLO} 2.6.3~\cite{Arnold:2008rz,Arnold:2011wj,Figy:2003nv,DelDuca:2001eu,DelDuca:2001fn} 
including both the vector boson fusion and gluon fusion production mechanisms, before showering the resulting Les Houches event~\cite{Alwall:2006yp} files using \texttt{Pythia 6}~\cite{Sjostrand:2006za} with the Z2 tune~\cite{Skands:2010ak}.
Events are generated at $s=8$ and $s=14$ TeV with the cteq6ll PDF set \cite{Pumplin:2002vw}. The mixing angle ranges from $\alpha=0$ to $\alpha=1.5$ in steps of $\Delta \alpha=0.3$ for $m_{\rm{H}}=126$ GeV.  
For each value of $\alpha$ and for each initial state $\mathcal{O}(1M\!-\!4M)$ events are generated. At $s=8$ TeV generator level cuts are
$|\eta(H)|<2.5$, $|\eta(j)|<5.0$ for the two required jets, $p_{Tj}>20$~GeV, $\Delta R_{jj}>0.6$, $m_{jj}>200$~GeV, and $p_{T,H} > 70$~GeV. At $s=14$ TeV the cuts are
identical, except the $m_{jj}$ cut is instead raised to $m_{jj}>400$~GeV.
No cuts are made on $\Delta\eta_{jj,min}$ or $\eta_{j_1} \times \eta_{j_2}$ at the generator level.  For the gluon fusion process, the full top mass dependence is retained in the loop, while the bottom quark contribution is neglected.  In the Higgs decay to $\tau\tau$ the Higgs is treated as a \cpeven scalar, since in this study $\tau$ polarization plays no role.  This prescription also effects $\tau$ kinematics, but only at a negligible level suppressed by $\mathcal{O}(m_\tau/p_T)$. 

As demonstrated in the experimental papers~\cite{Chatrchyan:2014nva,ATLAS-CONF-2013-108} the dominant backgrounds for $h+2j$ production followed by $h\to \tau\tau$ are $Zjj$, $W+$ jets and to a lesser extent $t\bar t$. We generate events for these processes at 8~TeV and 14~TeV using \texttt{SHERPA}~2.0.0~\cite{Gleisberg:2008ta} with a similar series of selection cuts ($|\eta(\tau)|<2.5$, $p_{Tj}>20$~GeV and $\Delta R_{jj}>0.6$) to those described for the signal above. We consider the electroweak and QCD production of $Zjj$ separately. 
We do not take into account backgrounds arising from $h\to WW$ production, which only lead to small changes in the the $e\mu$ channel in our study. We do not generate any QCD multijet backgrounds, which are important for jets faking taus when both taus decay hadronically (see below).

We show in Table~\ref{tab:signalrates} the cross-sections at parton level for the signal as a function of the mixing angle $\alpha$ for both the gluon fusion and vector boson fusion channels at 8~TeV (left) and 14~TeV (right). We observe that the WBF contribution decreases with increasing mixing angle $\alpha$ as expected, while the contribution from the gluon fusion component increases.

\begin{table}[h]
\vspace{5mm}
\begin{tabular}{|c||c|c||c|c|c|} \hline
$\alpha$ & 8~TeV GF cross-section (fb) & 8~TeV WBF cross-section (fb) & 14~TeV GF cross-section (fb) & 14~TeV WBF cross-section (fb)\\ \hline
0.00 & 250 & 467  & 1141 & 1481  \\ \hline
0.30 & 278 & 426 & 1268 & 1351 \\ \hline
0.60 & 352 & 318 & 1606 & 1009 \\ \hline
0.90 & 447 & 181 & 2038 & 572  \\ \hline
1.20 & 529 & 61 & 2411 & 194 \\ \hline
\end{tabular}
\caption{The gluon fusion and weak boson fusion signal cross-sections at the generator level before event selection and Higgs decay for 8~TeV (left) and 14~TeV (right).}
\label{tab:signalrates}
\end{table}

\subsection{Simulation Details}

We select four different final states with which to perform our analysis, classified by the $\tau$ decay channel. There is the fully hadronic di-$\tau_h$ case and the semi-leptonic and leptonic cases $e\tau_h$, $\mu \tau_h$ and $e \mu$. The initial selection cuts we apply to these final states are shown in Table~\ref{tab:tauselect}. The selection is intended to closely mimic both the CMS and ATLAS di-$\tau$ analysis.  The one missing background from the simulation is the QCD multijet background where a jet imitates a lepton or fake $\tau_h$. This background is particularly important in the di-$\tau_h$ final state. We assume that the QCD contribution is flat and uniformly covers the full phase space of the selected region. This is consistent with the results of~\cite{Chatrchyan:2014nva}. We set the normalization by considering the differential $m_{jj}$ cross-section from QCD, extrapolating this to the Z mass, and multiplying by the fake rate for a jet to fake a tau at 50~GeV. Following the selection, using the 8~TeV samples the yields are found to be comparable to both existing CMS and ATLAS results at the 10\% level.

\begin{table*}[pt!]
\begin{tabular}{|c|c|c|c|c|} \hline
                            & $\tau_h\tau_h$                                                & $\mu\tau_h$                                                                                 &  $e\tau_h$                                                              &  $e\mu$                                                                                       \\ \hline
lepton selection  & $p_{T}^{\tau} > 45$\GeV                                 & \begin{tabular}{@{}c@{}} $p_{T}^{\mu} >$ 20\GeV \\ $p_{T}^{\tau} > 30$\GeV  \end{tabular}                 & \begin{tabular}{@{}c@{}}  $p_{T}^{e} >$ 25\GeV \\ $p_{T}^{\tau} > 30$\GeV  \end{tabular}     &  \begin{tabular}{@{}c@{}}  $p_{T}^\mathrm{lead} >$ 20\GeV \\ $p_{T}^\mathrm{trail} > 10$\GeV          \end{tabular}           \\ \hline
kinematic selection  &  $p_{T}^{H} > 100$\GeV                             & $m_{T}^{\mu} < 30$\GeV                                                              & $m_{T}^{e} < 30$\GeV                                            & b-tag veto  with $p_{T}^{b} > 20$ GeV                                        \\ \hline
loose jet selection  & \begin{tabular}{@{}c@{}} $m_{jj} > 500$\GeV \\ $|\Delta\eta_{jj}|>$3.5 \end{tabular}  & \begin{tabular}{@{}c@{}} $m_{jj} > 500$\GeV \\ $|\Delta\eta_{jj}|>$3.5 \end{tabular}                              & \begin{tabular}{@{}c@{}} $m_{jj} > 500$\GeV \\ $|\Delta\eta_{jj}|>$3.5 \end{tabular}              & \begin{tabular}{@{}c@{}} $m_{jj} > 500$\GeV \\ $|\Delta\eta_{jj}|>$3.5 \end{tabular}                       \\ \hline
tight jet selection    &   & \begin{tabular}{@{}c@{} } $m_{jj} > 700$\GeV \\ $|\Delta\eta_{jj}|>4.5$ \\ $p_{T}^{H} > 100$\GeV \end{tabular} & \begin{tabular}{@{}c@{}} $m_{jj} > 700$\GeV \\ $|\Delta\eta_{jj}|>$4.5 \\ $p_{T}^{H} > 100$\GeV \end{tabular} & \begin{tabular}{@{}c@{}} $m_{jj} > 700$\GeV \\ $|\Delta\eta_{jj}|>$4.5 \\ $p_{T}^{H} > 100$\GeV \end{tabular} \\ \hline
\end{tabular}
\caption{Kinematic selection and jet selection for the four different channels ($\tau_h \tau_h$, $\mu \tau_h$, $e\tau_h$ and $e\mu$) used for our  di-$\tau$ analysis. The di-jet selection includes both exclusive loose and tight categories for all the channels apart from the $\tau_{h}\tau_{h}$ channel. }
\label{tab:tauselect}
\end{table*}

To emulate the performance of the detectors all reconstructed physics objects are smeared by a standard set of resolution functions. For the muons,
electrons, and $\tau_h$, resolution functions with widths of 2\GeV, 3\GeV, and 4\GeV~are used.
For the jets a series of resolution functions binned in $\eta$ is used. The parameterizations for these are taken from \cite{CMS:2011esa,Aad:2012ag}. 

The smearing is parameterized in the unclustered $\pt$ and smeared separately for the parallel and perpendicular components of the unclustered energy with respect to the Higgs $\pt$.  To simulate the instance of fake $\tau_h$ being  produced from a jet, the jet having the smallest energy in an annulus about the jet axis  of $0.1<\Delta R< 0.4$ and having a  $\pt > 20$\GeV ~is selected and deemed to be the fake $\tau_h$. Provided a fake $\tau_h$ candidate exists, the event is then reweighted as a function of the $\pt$ of the $\tau$ using the fake rates reported by CMS~\cite{CMS:xxa}. Finally, for events in the $e\tau_h$ channel a non-negligable fake background results from the instance where one electron is reconstructed as a fake $\tau_h$ candidate. To simulate this number we take the fake rate to be  roughly consistent with the tight working points for both the ATLAS and CMS anti-electron vetoes~\cite{Chatrchyan:2014nva,ATLAS-CONF-2013-108}. 
For the lepton efficiencies a flat efficiency corresponding to 90\% is taken for the muons, 80\% for the electrons and 60\% for the taus. These numbers take into account both the expected trigger and identification efficiencies for the leptons after a typical $e/\mu/\tau_h$ selection. For the $\tau_h$ efficiency in the case where an anti-electron veto is applied the corresponding $\tau_h$ efficiency is scaled down by an additional 10\%. For the $e\mu$  channel the efficiencies are scaled up by 5\% in electrons and muons, corresponding to the improved trigger efficiency in these cases.

The uncertainties applied in the extraction of the signal closely resemble the current LHC analyses. For each background, an uncorrelated normalization uncertainty of 10-50\% is applied in each category. The variation in the uncertainty is dependent on whether a real or fake $\tau_{h}$ is present. Additional correlated normalization uncertainties of 1-5\% are also applied reflecting the effects of lepton efficiencies, jet scale, and luminosity. No lepton energy scale uncertainty is applied, since this is well constrained from other categories in the LHC analyses. Regarding the theoretical uncertainties on the Higgs yields, the cross section uncertainties from the Higgs working group are applied~\cite{Heinemeyer:2013tqa}, along with an additional uncorrelated 25\% uncertainty on the overall gluon fusion yield to reflect  the current knowledge of di-jet production in gluon fusion. For projections with an integrated luminosity $>20~\ifb$, this uncertainty is reduced to 10\%, reflecting expected theoretical improvements in the signal yield calculation. As with the current LHC analyses, these systematic uncertainties are added to the signal extraction, separately floating each uncertainty under a gaussian prior whose width is specified by the systematic uncertainty.

 As with the CMS and ATLAS analyses, extraction of the signal relies on exploiting  full knowledge of the $\tau$ decays to improve the mass separation of the signal from the largest background, Z$\rightarrow\tau_h\tau_h$. Such a scenario benefits greatly from incorporating knowledge of the $\tau$ decay matrix elements into the kinematic reconstruction of the di-$\tau$ mass.  To perform this mass reconstruction, we developed a di-$\tau_h$ mass reconstruction that computes a weighted likelihood of the di-$\tau_h$ mass on an event-by-event basis by randomly sampling the allowed neutrino kinematics from the leading order matrix elements and weighting each event by the consistency with the observed missing transverse energy (MET), using the full MET resolution covariance matrix. This mass reconstruction is very similar to the MCT approach used by ATLAS and the SVFit mass approach used by CMS~\cite{Chatrchyan:2014nva,Elagin:2010aw}.
As a final cross check, our simulation was checked against the current CMS di-$\tau$ analysis and gave yields, shapes and results similar to those reported in their paper~\cite{Chatrchyan:2014nva}.

\subsection{Observable distributions} 

\begin{figure*}[pt!]
\includegraphics[height=0.35\textwidth,width=0.35\textwidth]{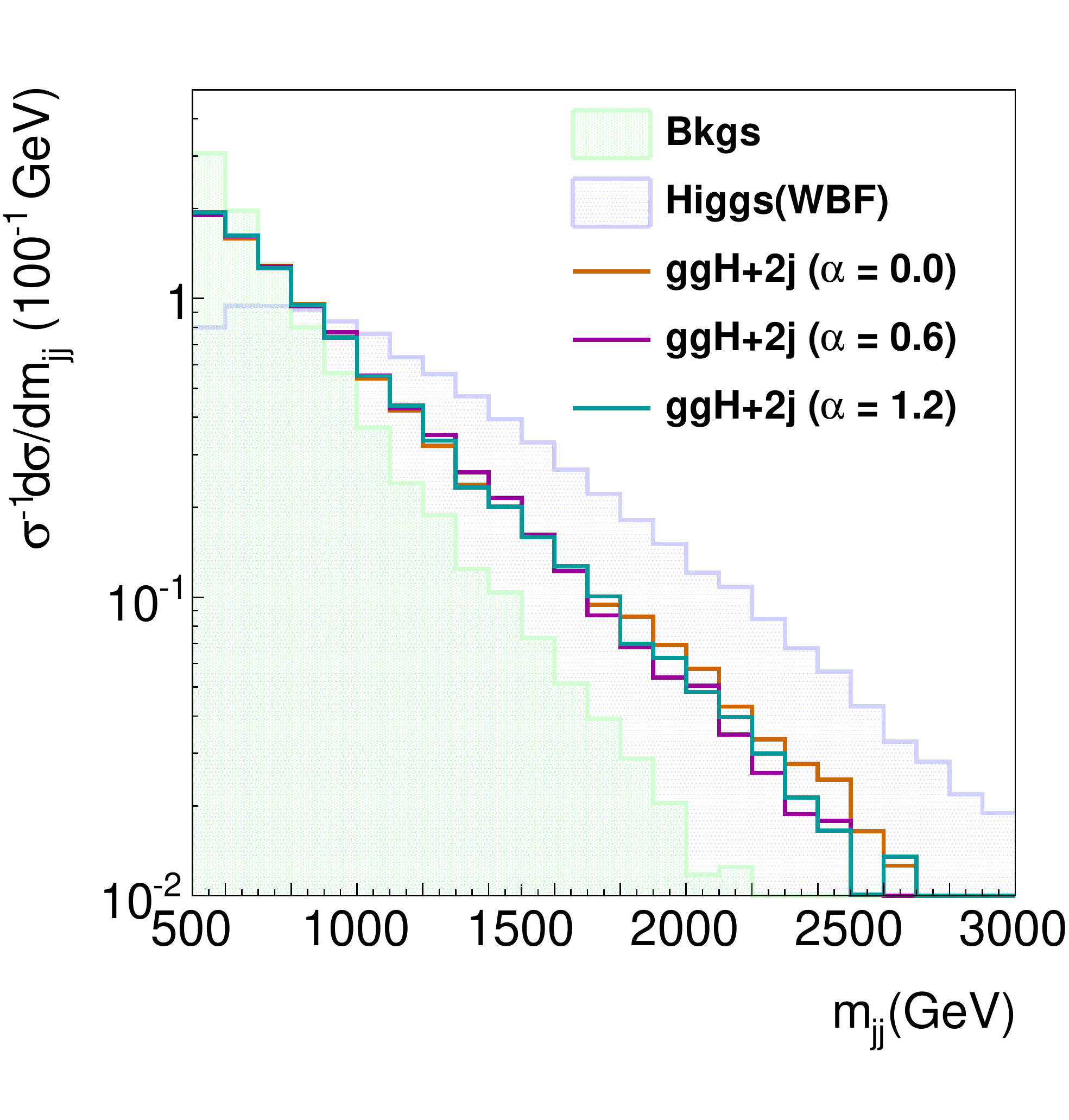}\hspace{1cm}
\includegraphics[height=0.35\textwidth,width=0.35\textwidth]{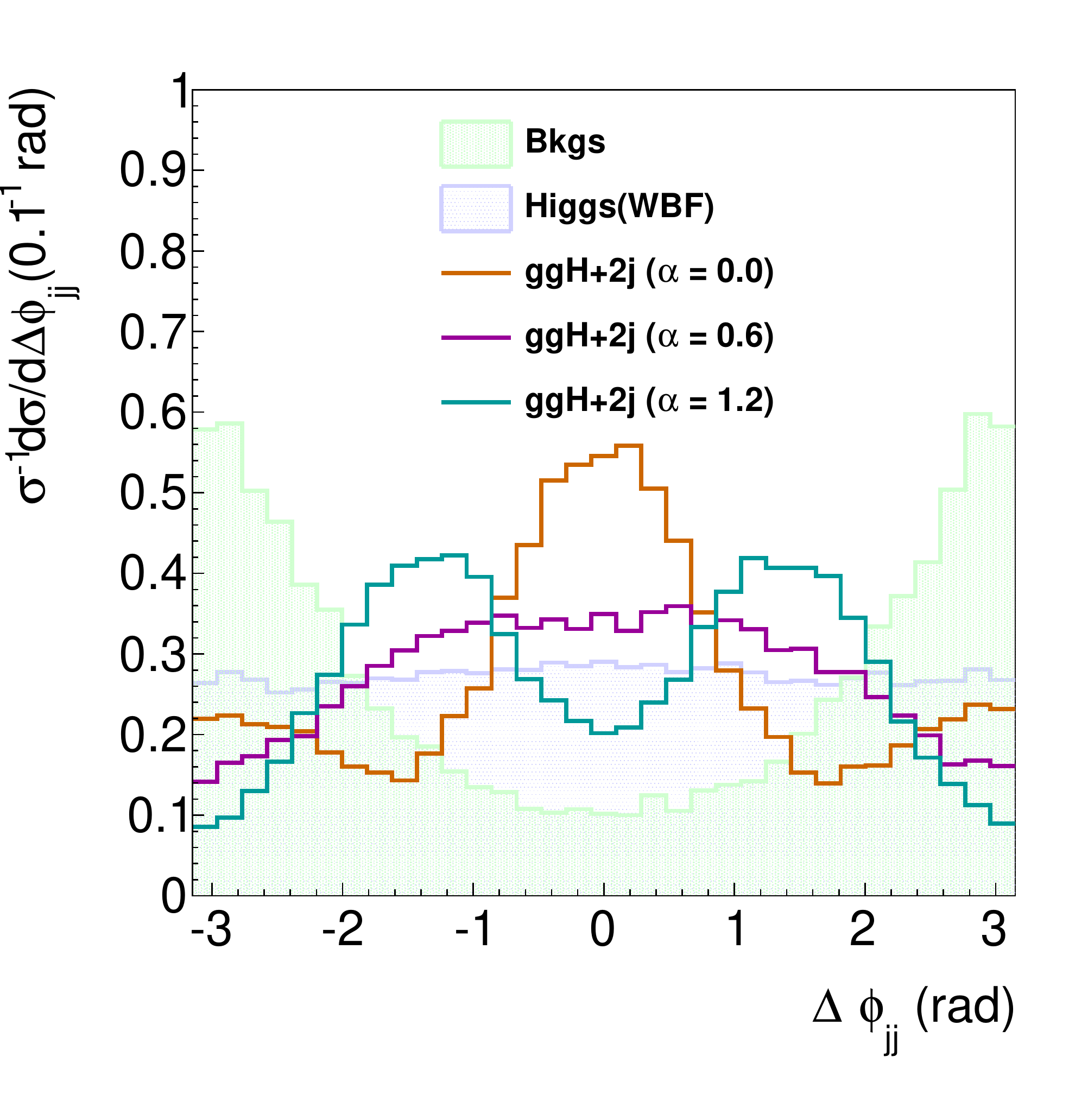}\\
\includegraphics[height=0.35\textwidth,width=0.35\textwidth]{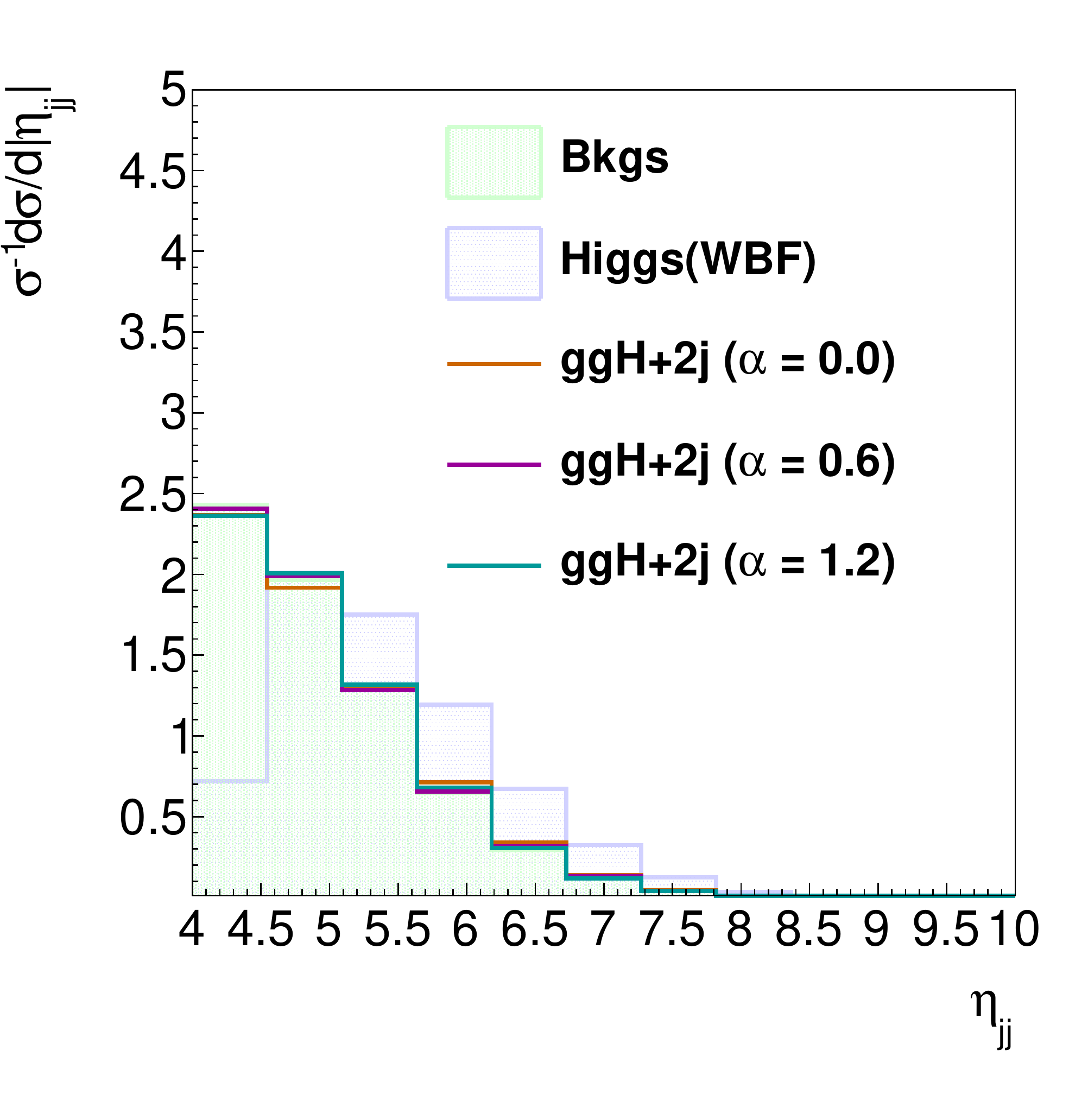}\hspace{1.0cm}
\includegraphics[height=0.35\textwidth,width=0.35\textwidth]{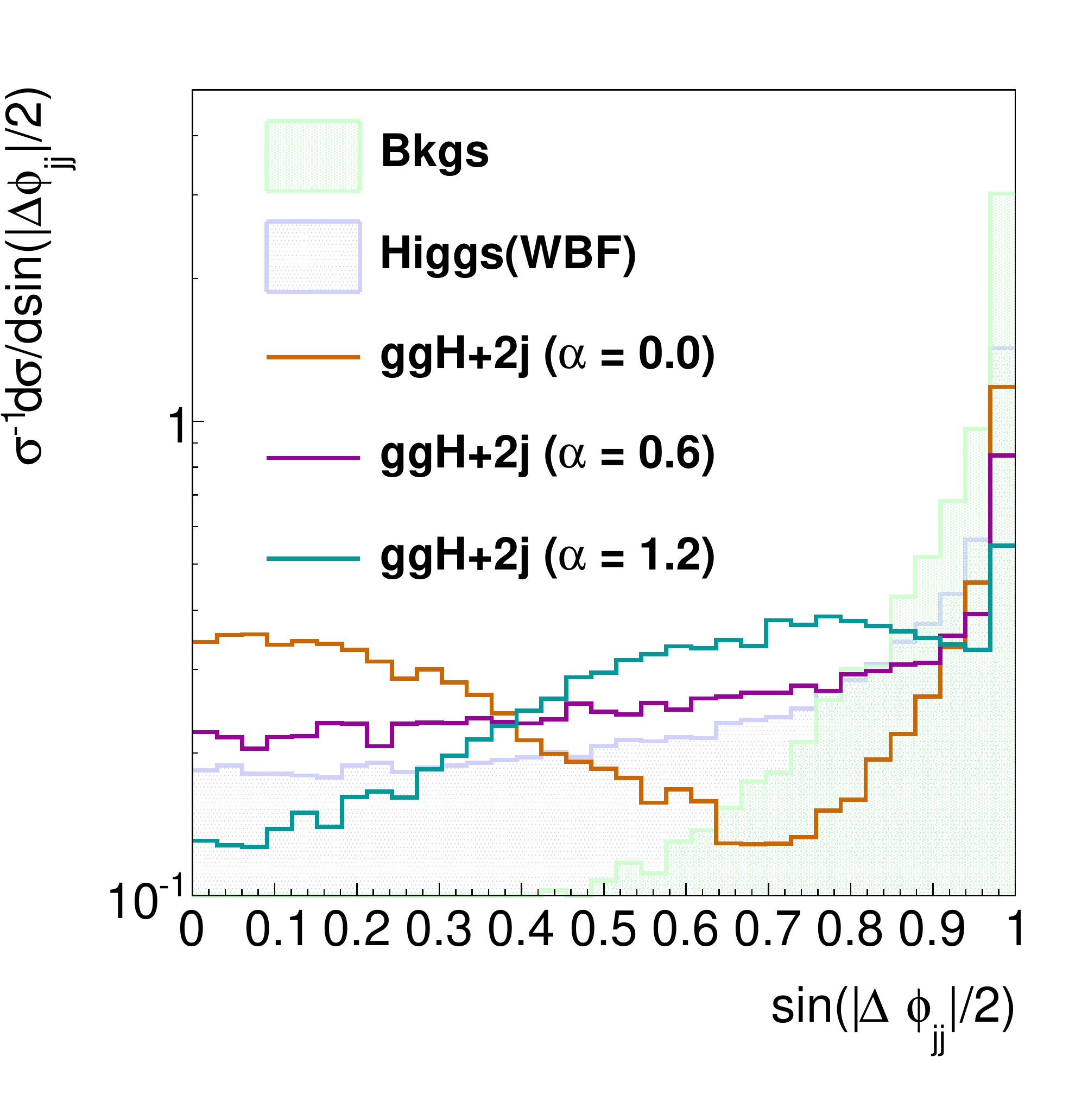}
\caption{\label{fig:sigplots} Observable distributions for the signal and background. From the top-left and proceeding clockwise: $m_{jj}$, $\Delta\phi_{jj}$, sin($|\Delta\phi_{jj}|/2$) and $\Delta \eta_{jj}$. For each figure the yields are normalized to the expected yields at 8 TeV for the gluon fusion
channel at 20 fb$^{-1}$ with $\alpha=0$. Samples have been passed through the detector pseudo-simulation and subjected to the full selection on all channels. The loose WBF selection and the additional category selections are applied in all cases. }
\end{figure*}

We show in Fig.~\ref{fig:sigplots} starting from the top-left and working clockwise the distributions for the invariant mass $m_{jj}$ of the two tagging jets, the azimuthal angle difference $\Delta \phi_{jj}$ between the tagging jets, the rapidity difference $\Delta \eta_{jj}$ between the jets and finally the discriminating variable $\sin\left( |\Delta\phi_{jj}|/2\right)$.  Each figure shows the total background contribution, along with that from WBF Higgs production for $\alpha=0$ and the GF signal component for $\alpha = 0, 0.6$ and $1.2$. The individual contributions are  normalized to the expected yields at 8~TeV for $20~\ifb$ for $\alpha=0$.
The variable showing the largest sensitivity to the mixing angle is the azimuthal angle between the two tagging jets, $\Delta\phi_{jj}= \phi_{y>0} - \phi_{y<0}$, which has long been known to provide a good handle for discriminating Higgs \cp~properties~\cite{DelDuca:2006hk,Klamke:2007cu}.  In addition some small dependence on $\alpha$ can also be observed at large values of the dijet invariant mass $m_{jj}$.  All the distributions we show are for events that have been showered and smeared using our detector pseudo-simulation. We have also investigated the $\pt$ distribution of the leading jet, which shows some limited sensitivity to $\alpha$ near the peak of the distribution.

As a cross check of the possible performance gain that can be had by utilizing other observables we have applied a boosted decision tree (BDT) that was trained to discriminate a fully simulated gluon fusion sample with $\alpha=1.2$ from one with $\alpha=0$. To train this decision tree, we used 18 observables obtained from the pseudo-simulation. These include the two leading jet $\eta$'s and $p_T$'s, the 3-vectors for the visible components of the $\tau$ decays, the kinematically fitted mass $m_{\tau\tau}$, the Higgs $p_T$ constructed from the MET and the visible decay products, the MET, the transverse mass of either lepton combined with the MET, and the  $m_{jj}$, $\Delta \eta_{jj}$, and $\Delta \phi_{jj}$ variables.   The training was performed separately for each channel, so as to improve the individual performance of each observable. The performance gain of these variables with respect to $\sin\left(|\Delta\phi_{jj}|/2\right)$ is shown in Fig.~\ref{fig:roccurves} for both 8 and 14~TeV.

As part of the optimization studies for the WBF selection, a BDT was used to train both the WBF and gluon fusion signals against a weighted sum of all the backgrounds using the same variables as described in the previous paragraph. After the optimization, only marginal gains were found beyond the addition of four main variables, $m_{jj}$, $|\Delta\eta_{jj}|$, the di-$\tau$ mass $m_{\tau\tau}$, and $\Delta\phi_{jj}$. The addition of $\Delta\phi_{jj}$, in particular, brought a performance improvement  of 20\% in the WBF sensitivity. In both CMS and ATLAS,
this variable had been used minimally, so as to avoid spin sensitivity and to avoid complications resulting from theoretical modeling of the second jet in gluon fusion.  Once $\Delta\phi_{jj}$ was added, it was further found that a category-based analysis binning in mass, $\Delta\phi_{jj}$, $m_{jj}$ and $\Delta\eta_{jj}$ performed as well  as a BDT trained on the full set of observables.

\begin{figure*}[pt!]
\includegraphics[height=0.35\textwidth,width=0.35\textwidth]{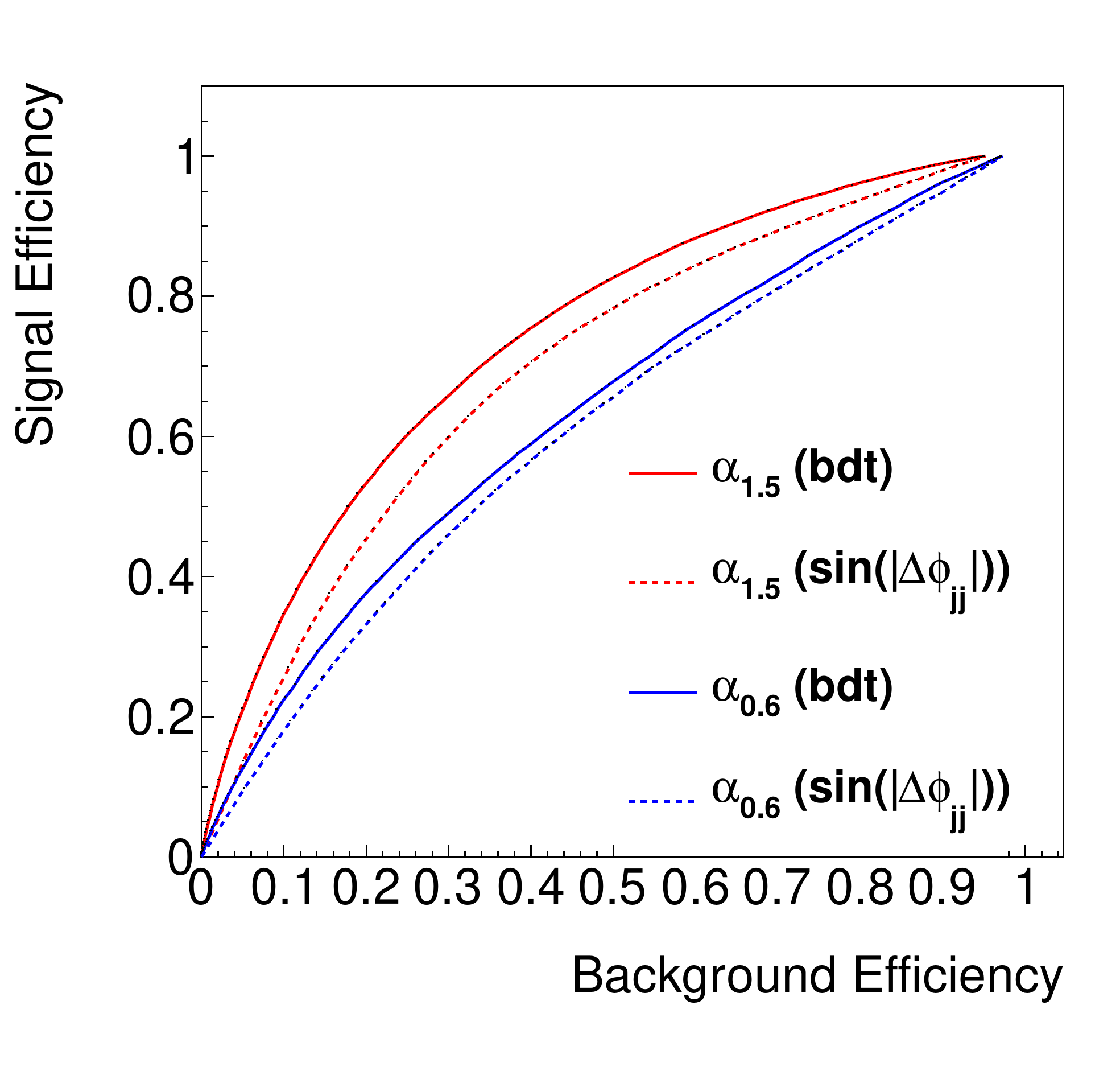}\hspace{1cm}
\includegraphics[height=0.35\textwidth,width=0.35\textwidth]{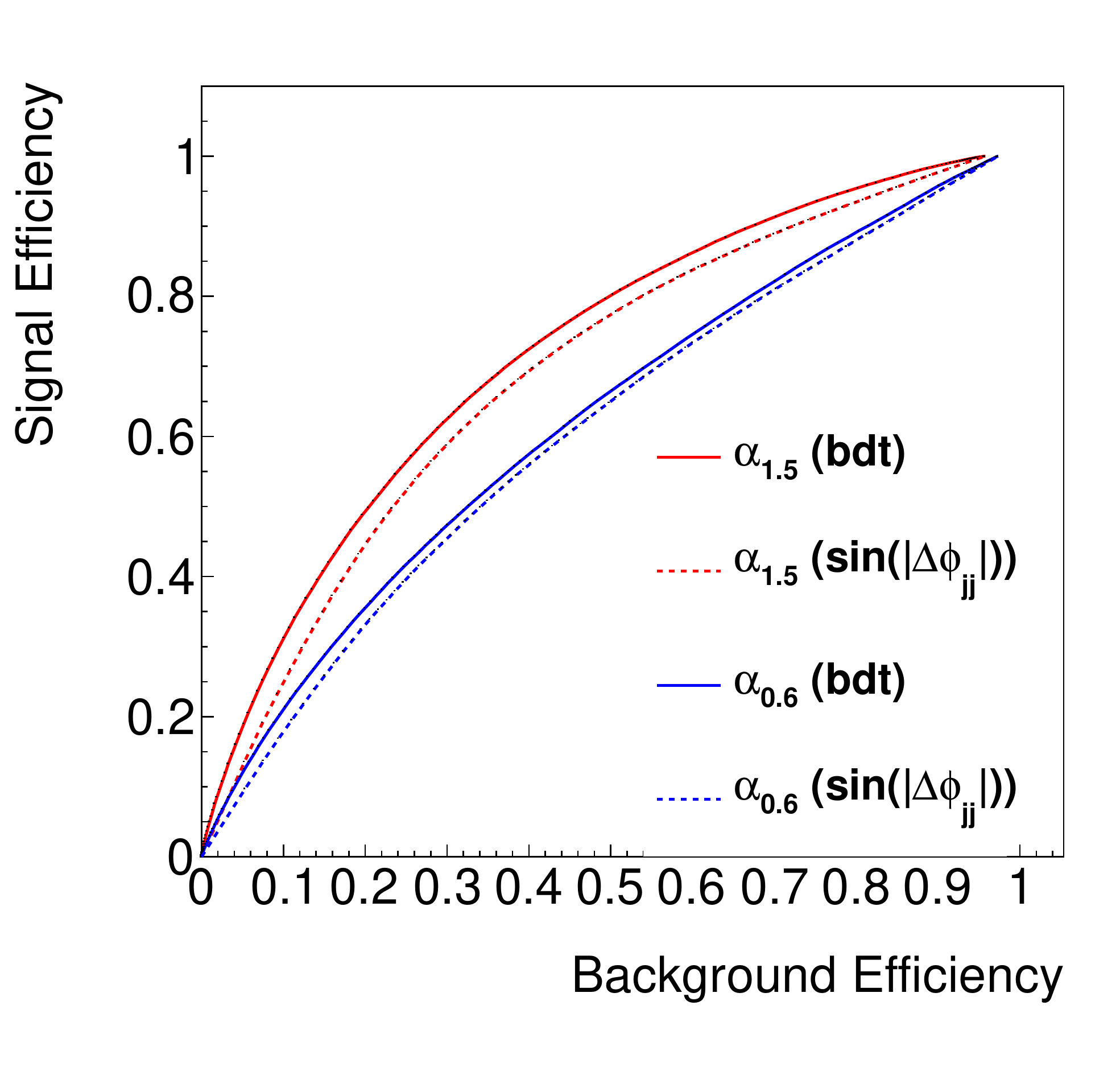}
\caption{\label{fig:roccurves} RoC curves for our boosted decision tree for 8~TeV (left) and 14~TeV (right). The red lines are for $\alpha=1.5$ and the blue lines for $\alpha=0.6$. The dashed curves show the results only including the $\sin\left(|\Delta\phi_{jj}| \right)$ variable, and the solid curves those for the full BDT with all 18 observables included, as described in the text.}
\end{figure*}

\subsection{$\Delta \phi_{jj}$ analytics} 

We now briefly discuss the the $\Delta \phi_{jj}$ dependence of the two different production mechanisms. To begin, consider the gluon 
fusion process (specifically $gg \rightarrow Hgg$) in the $m_t \to \infty$ limit. Apart from the strength of the coupling constants in Eqn.~\ref{eqn:lhgg}, the only difference between $\alpha=0$ and $\alpha=\pi/2$ is to be found in the form of the helicity conserving amplitudes \cite{Dawson:1991au,Kauffman:1996ix,Kauffman:1998yg,Kauffman:1999ie}. We note that the helicity violating amplitudes do exhibit $\Delta \phi_{jj}$ dependence, but the resulting terms are independent of the mixing angle $\alpha$ (apart from the strength of the coupling constants). Consider a
final state configuration in which the Higgs is central ($y_H=0$) and the two jets have opposite rapidities ($y_j \equiv y_{j1} = -y_{j2}$). Given our selection cuts described above, this represents a typical final state.
As a further simplification let the lab frame and the center-of-mass (CM) frame be identical so that the initial state gluons have equal and opposite 3-momenta, $\vec{p} = \pm \tfrac{1}{2}E_{\rm{CM}} \hat{z}$, where $\hat{z}$ is the direction along the beam axis.
In the limit where the final state jets have large rapidities, one finds that the helicity conserving squared matrix element for \cpeven ($+$ sign) and 
 \cpodd ($-$ sign) is given by (omitting coupling constants and other numerical prefactors)
\begin{equation}
\label{eqn:msqgf}
|\mathcal{M}|^2_{\rm{GF}\pm} \sim  \exp(4y_j)\{A\pm B \cos(2\Delta \phi_{jj})\}
\end{equation}
where
\begin{equation}
A=\xi^4+\xi^{-4} +\frac{1}{2}(\xi^5+\xi^{-3}) \qquad \rm{and} \qquad
B=2+\xi^2
 \qquad \rm{with} \qquad \xi \equiv \frac{E_{\rm{CM}}}{E_{\rm{CM}}-m_h}
\end{equation}
so that in the limit where $m_h \ll E_{\rm{CM}}$ we have
\begin{equation}
|\mathcal{M}|^2_{\rm{GF}\pm} \sim \exp(4y_j)\{3\pm 3\cos(2\Delta \phi_{jj})\}
\end{equation}
In the case of the WBF production mechanism ref.~\cite{Plehn:2001nj} argued that the matrix element squared in the limit of forward
jets is approximately given by
\begin{equation}
\label{eqn:msqwbf}
|\mathcal{M}|^2_{\rm{WBF}} \sim \hat{s}\;\! m_{jj}^2
\end{equation}
 where $\hat{s}$ is the partonic center-of-mass energy, which results in an essentially flat distribution in $\Delta \phi_{jj}$.  Note however that
the dimension five operators that we assume to be negligible and omit ($h V^{\mu\nu} V_{\mu\nu}$ and $h V^{\mu\nu} \widetilde V_{\mu\nu}$) lead to
a non-trivial $\Delta \phi_{jj}$ dependence.

The dependence of the gluon fusion and WBF production mechanisms on $\Delta \phi_{jj}$ can be seen explicitly in Fig.~\ref{fig:sigplots}. The form
of the distributions follows the expectations from Eqn.~\ref{eqn:msqgf} and Eqn.~\ref{eqn:msqwbf}.  The fact that the approximations leading to
these matrix elements are quite good makes it clear why $\Delta \phi_{jj}$ by itself is nearly optimal as a discriminating observable between the
\cpeven and \cpodd case. In principle, we can include in Eqn.~\ref{eqn:msqgf} the next term in the series in $\exp(y_j)$.  Doing so upsets
the factorized form $|\mathcal{M}|^2_{\rm{GF}\pm} \sim f(y_j)g(\Delta \phi_{jj})$. 
In particular, the next term in the series, which is proportional to $\exp(2y_j)$, includes subterms with $\cos(n\Delta \phi_{jj})$ for $n=1,2,3$ and breaks the degeneracy $|\mathcal{M}|^2_{\rm{GF}\pm}(\Delta \phi_{jj})=|\mathcal{M}|^2_{\rm{GF}\pm}(\Delta \phi_{jj}\pm \pi)$ in Eqn.~\ref{eqn:msqgf}, reflecting the observed behavior in Fig.~\ref{fig:sigplots}. 
This clarifies why the BDT has an edge in discriminatory power. However, since the correction due to the next term in $\exp(y_j)$ is small for the phase space region of interest, the BDT exhibits only marginally better discriminatory power
than $\Delta \phi_{jj}$ by itself (as is evidenced in Fig.~\ref{fig:roccurves}).


\section{Estimated Limits on \cp~Properties}
\label{sec:results}

We now discuss our results. In Fig.~\ref{fig:moneyplots} we show the significances that can be achieved using the $20~\ifb$ of data from the 8~TeV run and projected limits for $50~\ifb$ of data at 14~TeV, corresponding to around two years of running.  Those results from the analyses marked with `Loose' were performed using the loose analysis cuts from Table~\ref{tab:tauselect}, while those marked `Tight' were performed with the tight analysis cuts from the same table, which forms a subset of the loose category. 

The dashed lines show the estimated significance of the total signal over the Standard Model backgrounds.  The dark yellow dashed line shows the results obtained doing a standard WBF-style analysis with loose cuts, achieving a significance of barely $2\sigma$ over the background. The purple dashed line shows our best approximation to the current CMS analysis~\cite{Chatrchyan:2014nva} with tighter cuts, while the upper three dashed lines either utilize the 18 variables described above (blue dashed) or use a simultaneous fit to both the di-tau invariant mass $m_{\tau\tau}$ and the discriminating variable $\sin \left(|\Delta\phi_{jj}|/2 \right)$ (green and maroon dashed, with loose and tight cuts respectively). 

\begin{figure*}[pt!]
\includegraphics[height=0.45\textwidth,width=0.45\textwidth]{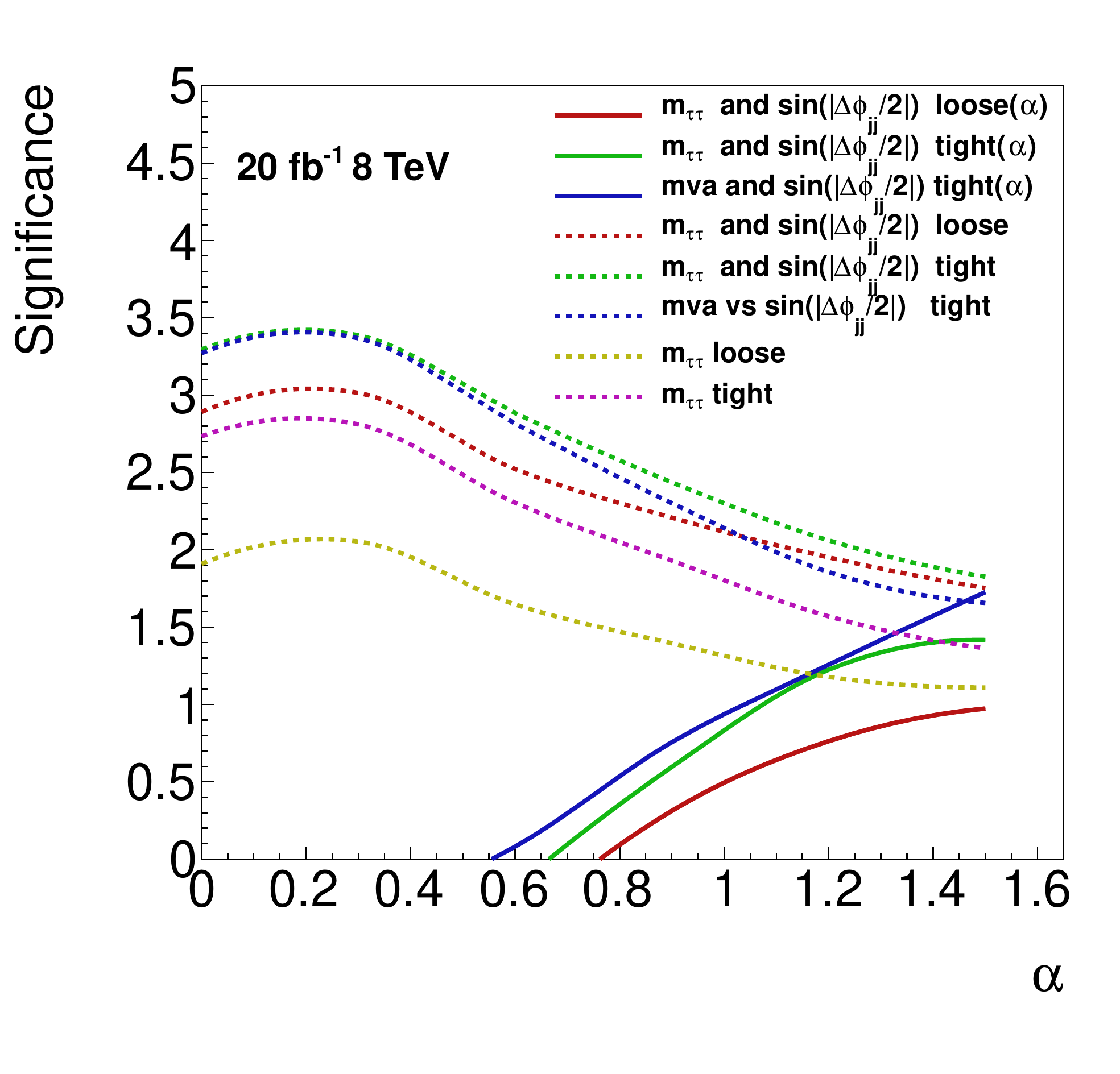}\hspace{0.5cm}
\includegraphics[height=0.45\textwidth,width=0.45\textwidth]{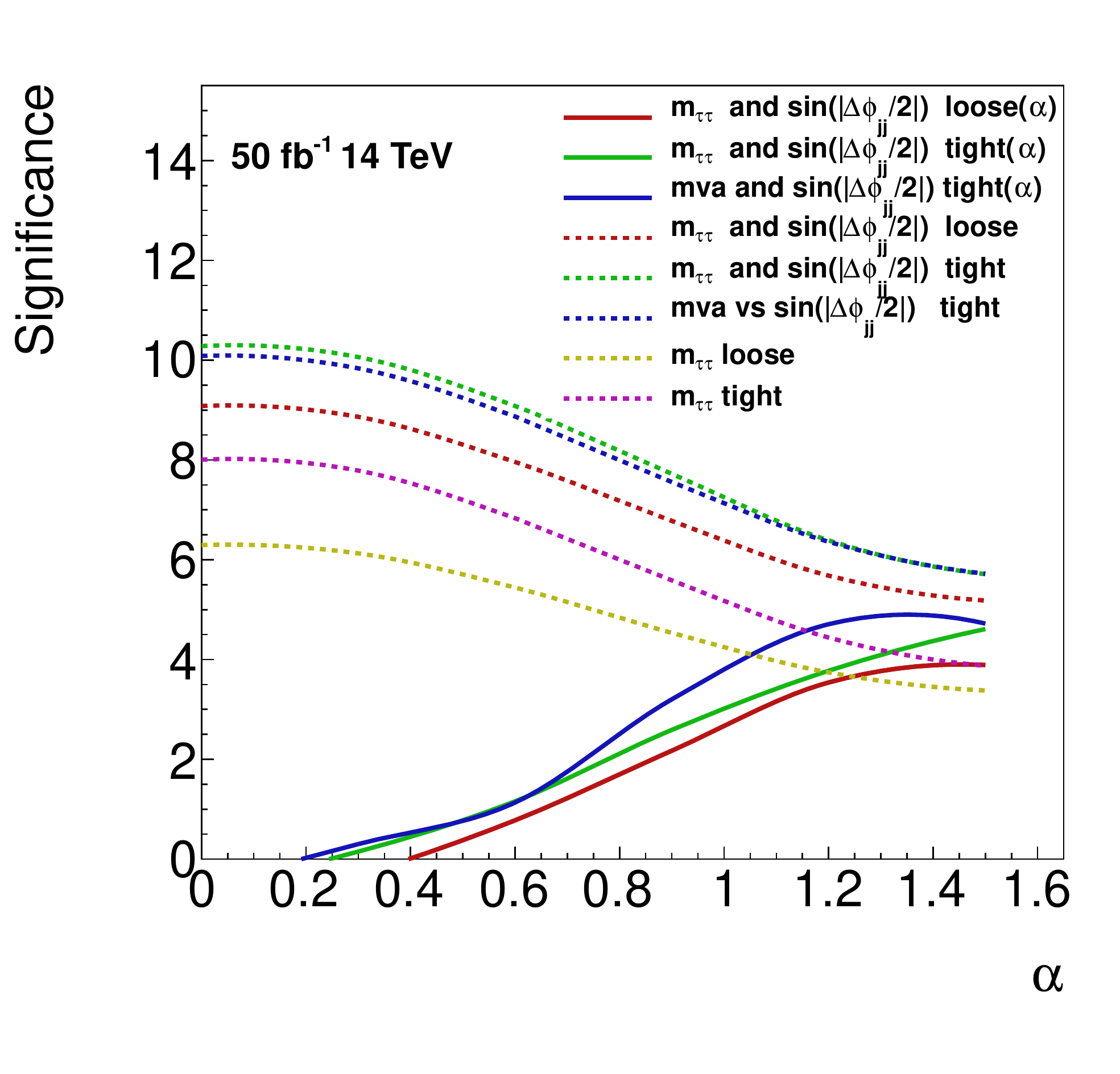}\\
\caption{\label{fig:moneyplots} Expected limits that can be achieved with our analysis using the $20~\ifb$ 8~TeV dataset (left) 
and using a $50~\ifb$ dataset at 14~TeV (right). The dashed lines show the estimated significance of the total signal over the Standard Model backgrounds and the solid lines show the exclusion significance computed using the $\mathrm{CL}_s$ method relative to the $\alpha=0$ case.  See text for details.}
\end{figure*}

The solid lines show the exclusion significance computed using the $\mathrm{CL}_s$ method~\cite{Read:2000ru} relative to the $\alpha=0$ case.  The maroon line again shows the results using the loose event selection and di-tau invariant mass and $\sin \left(|\Delta\phi_{jj}/2| \right)$, while the blue and green lines utilize the tight selection and (in the green case) the MVA. We observe from the left hand figure that with our best analysis a pure \cpodd Higgs corresponding to  $\alpha=\pi/2$ is already nearly ruled out at 95\% C.L. With $20~\ifb$ of luminosity at 14~TeV this should improve to $\alpha \leq 0.9$, while with $50~\ifb$ of luminosity it should improve further to $\alpha \leq 0.7$.

To further elucidate how the constraints on \cp-mixing will improve, in Fig.~\ref{fig:lumivsalpha} we show the expected exclusion limit on the mixing angle $\alpha$ as a function of the integrated luminosity at 14~TeV. This shows that the limit should improve to $\alpha \leq 0.3$ with $500~\ifb$.
As can be seen from the figure precision measurements of Higgs \cp~properties will benefit greatly from a high luminosity LHC run.

We note that the limits we have set can in principle be improved upon by including other techniques which are sensitive to the \cp~properties of the Higgs, such as including detailed information about the $\tau$ decay products as in~\cite{Bower:2002zx,Berge:2008wi}. Further discriminatory power between the gluon fusion and weak boson fusion production mechanisms could also be gained by using likelihood methods as in~\cite{Andersen:2012kn}. We are thus hopeful that it may be possible to improve upon our projections. With a similar analysis it may even be possible to extract information from the $h\to b\bar{b}$ decay.

\begin{figure*}[pt!]
\includegraphics[height=0.5\textwidth,width=0.5\textwidth]{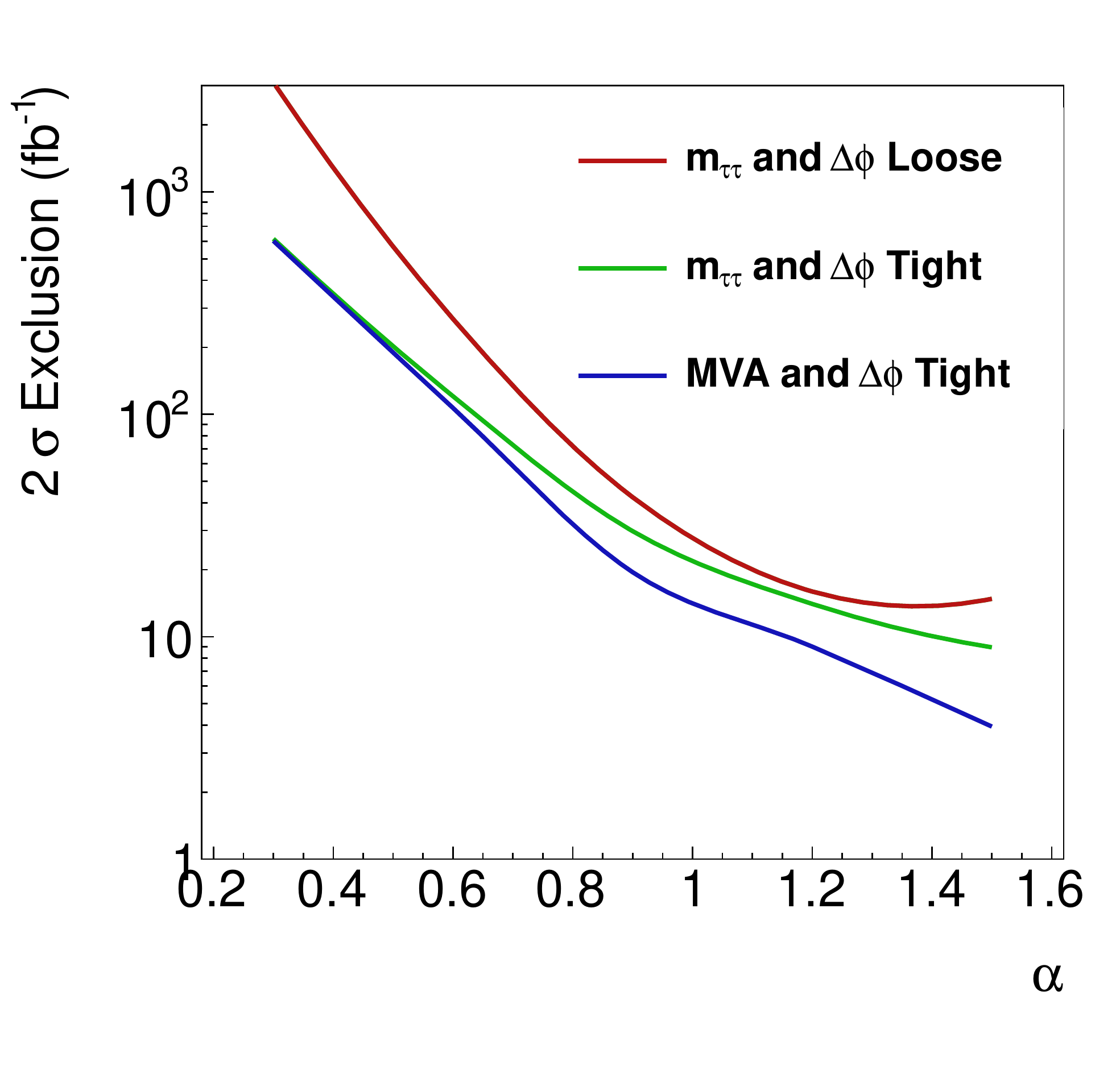}
\caption{\label{fig:lumivsalpha} The projected 95\% exclusion confidence limit on the mixing angle $\alpha$ that can be set as a function of the integrated luminosity at the 14~TeV run of the LHC. }
\end{figure*}

\section{Conclusions}
\label{sec:conclusion}

Measuring the detailed \cp~properties of the Higgs is one of the most important aspects of the precision Higgs program in the upcoming 14~TeV run at the LHC. Previous theoretical and current experimental analyses have focused on exploiting the Higgs couplings to massive vector bosons. However, the \cpodd couplings to $W$ and $Z$ are suppressed, so that analyses based on these couplings project out much of the physics of interest. Instead, we focus on Higgs interactions that have the same parametric strength for the \cpeven and odd Higgs components. This led us to consider Higgs production in association with two jets, followed by Higgs decay into a pair of $\tau$ leptons. Our analysis exploits the jet correlations in Higgs production, and is thus relatively independent of the \cp~nature of the $h\tau\tau$ coupling. Changes in the $h\tau\tau$ coupling will change the statistics, but not affect in any fundamental way our ability to set a limit on the \cp~mixing in this channel.

We have carried out a detailed simulation of the signal and backgrounds taking detector effects such as acceptances and fake rates into account and used a multivariate analysis to achieve excellent discriminating power in the mixing angle $\alpha$. We have presented estimates of the constraints that can be set using the current 8~TeV dataset, as well as 20 and $50~\ifb$ of data at 14~TeV, corresponding to approximately one and two years of running. We find that the 8~TeV dataset should be able to achieve nearly 95\% C.L. exclusion of a \cpodd Higgs relative to a \cpeven one. This should improve even further with the 14~TeV run such that $\alpha \geq 0.7 $ could be excluded with $50~\ifb$ and ${\alpha \geq 0.3 }$ with $500~\ifb$. By including other Higgs decay modes, e.g. $H \to \gamma \gamma$, the exclusion reach can be extended even further.

\section{Acknowledgments}
MJ would like to thank Franziska Schissler for assistance with VBFNLO.

\bibliography{ref}
\bibliographystyle{ArXiv}

\end{document}